\documentstyle[preprint,aps]{revtex}
\begin{document}
\draft
\newcommand{\be}{\begin{equation}}
\newcommand{\ee}{\end{equation}}
\newcommand{\bea}{\begin{eqnarray}}
\newcommand{\eea}{\end{eqnarray}}
\preprint{IOP-BBSR/95-}
\title{Fifty Years of the Exact solution of the Two-dimensional \\
Ising Model by Onsager}
\author{Somendra M. Bhattacharjee and Avinash Khare }
\address{Institute of Physics, Sachivalaya Marg,\\
Bhubaneswar 751 005, India.}
\date{\today}
\maketitle
\begin{abstract}
The exact solution of the two-dimensional Ising model by Onsager
in 1944 represents one of the landmarks in theoretical physics.
On the occassion of the fifty years of the exact solution, we
give a historical review of this model. After briefly discussing
the exact solution by Onsager, we point out some of the recent
developments in this field. The exact solution by Onsager has
inspired several developments in various other fields. Some of
these are also briefly mentioned.
\end{abstract}

\narrowtext
\section{INTRODUCTION}

It is usually said that every theoretical physicist must know
(ideally must be taught) the landmarks of theoretical physics.
Some of these are: $g-2$
calculation in Quantum Electrodynamics, Onsager's Exact
Solution of the Two-Dimensional Ising Model etc. It is our privilage
to briefly discuss  Onsager's exact solution of the Ising
model and its influence in the subsequent developments of
statistical mechanics \cite{bk}.  To appreciate as to why it is
regarded as one of the landmarks of theoretical physics, it is
worth remembering that for a long time it remained the first and
the only (mathematically rigorously) exactly solvable model
exhibiting phase transition. Its discovery completely changed
the course of developments in Statistical Mechanics and also
other areas of physics. Before Onsager's exact solution it was
not clear if the formalism of statistical mechanics can handle
phase transition.  The solution established beyond doubt that
phase transitions appear as singularities in the thermodynamic
functions and these functions need not have simple discontinuities as
hypothesized by Ehrenfest before.
Furhtermore, of all the
systems in statistical mechanics on which exact calculations
have been performed, the 2-dimensional Ising model is not only
the most thoroughly investigated but is also the
most profound. Its significance was instantly recognised. In
this context we would like to recall the letter written by Pauli
to Casimir immediately after the World War II. Casimir in his
letter had expressed his concern about being cut off for so long
from theoretical physics of allied countries. Pauli in his reply
said "nothing much of interest has happened except for Onsager's
exact solution of the Two-Dimensional Ising Model" \cite{obit}.

The plan of the article is as follows:In sec.II we first define
the Ising model and give an historical account of the
various approximate solutions developed leading to the exact
solution of Onsager \cite{nb}. We shall also briefly mention the
relevance of the model. A short sketch of the life and works of
Onsager are given in sec.III. In sec.IV we review the
exact solution of Onsager \cite{on} and its importance. Some of
the open, unsolved problems are also mentioned here.  It must be
made clear here that our aim is not really to discuss the exact
solution since there are several places where an excellent
account has already been given. In sec.V we discuss the various
developments in this field which have been directly inspired or
influenced by Onsager's work.

\section{THE MODEL AND ITS BRIEF HISTORY}

The Ising model was infact first written down not by Ising but
by his thesis supervisor W. Lenz \cite{le} in 1920 who was then
working in Rostalk University in Germany. It is somewhat
unfortunate that the physics community has given him no credit
for this work.  This is perhaps because even though Lenz
introduced the model, he did not do any calculations using this
model. It is perhaps worth recalling that Lenz is instead well
known for his work on Runge-Lenz vector in the context of the
accidental degeneracy in the hydrogen atom problem. Lenz moved
to Hamburg as professor in 1921. One of his first Ph.D. students
was Ernest Ising who was born on May 10, 1900. In late 1922,
Lenz asked Ising to study his model and the phenomena of
ferromagnetism. Ising studied the model and found its exact
solution in one dimension and showed that there is no phase
transition from para to ferromagnetism \cite{is}. Since then the
model is known as Ising model.  Let us now briefly state the
Ising model.

{\bf The Model:} It is a lattice model. Consider a d-dimensional
periodic lattice (d = 1,2,3,...) having an array of N fixed
points. The lattice may be of any type.  For example, one could
have 3-dimensional cubic or hexagonal lattice. However, we shall
be mostly considering the two dimensional square lattice since
it is for this that Onsager obtained his exact solution. With
each lattice site is associated a spin variable $S_i$ (i =
1,2,...,N) which is a number being +1 or -1. If it is 1 or -1 we
call it spin up or spin down. Clearly, a given set of numbers
\{$S_i$\} specify a configuration of the whole system and that
there are in all $2^N$ different configurations.The energy of
the system in a given configuration {$S_i$} is defined to be
\be \label{eq1}
E \{ S_i\} = - \sum_{<ij>} J_{ij} S_i S_j - B \sum^N_{i=1}
S_i,
\ee
where B is the external magnetic field. Usually one assumes that
the interaction is isotropic (i.e. $J_{ij} = J > 0$ ) and
that there is only nearest neighbour interaction. For example,
for a 2-dimensional square lattice the number of nearest
neighbours is 4. Given the energy of the system, the goal is to
compute the canonical partition function ($\beta=(kT)^{-1}$,
where $k$ is the Boltzmann consatnt and $T$ is the temperature)
\be \label{eq2}
Q{(B,T)} = \sum_{S_1}\sum_{S_2} .... \sum_{S_N} e^{-\beta
E \{S_i\}},
\ee
and hence the thermodynamic properties, and to see if the model
exhibits any phase transition at a finite nonzero temperature. In
other words, in magnetic systems, each molecule has a spin that
can orient up or down relative to the direction of the applied
external magnetic field B. The question one would like to ask is
if this model could lead to spontaneous magnetisation i.e. if
below a certain critical temperature $T_c$, $M  (B=0,T)$ which is
essentially same as the order parameter $\overline L (B=0,T)$, is
nonzero or not. Here the magnetization $M  (B,T)$ is
related to the partition function by
\be \label{eq3}
M  (B,T) = + {\partial\over\partial B} [ \ln Q  (B, T)].
\ee
It is worth pointing out here that the Ising model equally well
represents a model for (i) Binary Alloys and (ii) Lattice Gas.
Binary alloys are mixtures of two types of molecules (for example
$\beta-$brass has a cubic structure made out of Zn and Cu atoms)
and the question is if below a certain temperature $T_c$, there
is a phase transition with atoms of the same type clustering
together.  On the other hand, in lattice gas models one
considers a mixture of molecules and holes(i.e. empty spaces) on a
latice and
the question is if below $T_c$, there is a condensation of molecules
into one region of space and holes in the rest of the lattice?
Yang and Lee \cite{yl} discussed this model and gave a detailed
mapping between it and the Ising model. It may be noted here
that the lattice gas models are of relevance in the context of
gas-liquid and liquid-solid transitions. We thus see that the
study of the Ising model is of relevance in a number of phenomena.
Infact for $J<0$, this model could very well
represent (i) antiferromagnetic ordering, (ii) superlattice
structure in an alloy, and (iii) a solid-like arrangement of
molecules with repulsive forces.

A point in order at this stage. Many people tend to dismiss the
Ising model as over-simplified representation of intermolecular
forces. However the point to note is that the essential features
of the cooperative phenomena(i.e. long range order), specially
near $T_c$, do not depend on the details of the intermolecular
forces but on the mechanism for propagation of long range order
and Ising model offers much hope for an accurate study of this
mechanism.

{\bf Historical Developments:} As mentioned above, Ising was
able to obtain the exact solution of the model in 1-dimension
\cite{is} and show that there
is no phase transition at $T\neq 0$. He then gave some arguments
and erroneously concluded that even in two or three dimensions
this model will not exhibit any phase transition. This dissuaded
several people from working on this model and prompted
Heisenberg \cite{he} to introduce more complicated vector
interaction between spins known popularly as Heisenberg
ferromagnetic interaction.  It is interesting to note that Ising
seriously believed in his conclusions and felt very frustrated
that the model has no usefulness and  gave up physics
research! Being a Jew, he was prosecuted in Germany and was
dismissed from his job in 1933. He managed to run away from
Germany and lived in Luxemberg from 1939 to 1947. During all
this time he had no contact with physics and lost track of
future developments about the model. Only in 1947 when he went
to U.S.A. to teach did he realize that his name has become
immortal because of the work done by other people!

Almost a decade after Ising's work, Bragg and Williams \cite{bw}
studied the model in the mean field approximation. They were
inspired by an earlier work of Gorsky \cite{go}. The essential
assumption here is that the energy of an individual atom in any
configuration is determined by the average degree of order
prevailing in the entire system rather than by the fluctuating
configurations of the neighboring atoms.  Clearly, this
approximation is exact in the limit $d \rightarrow \infty$. One
of the serious criticism of this approximation is that it is
independent of the dimension d. This approximation predicts
a phase transition in the Ising model at finite T for any d which
is clearly wrong for d = 1 as one knows from Ising's exact
calculation.

Soon afterwords, Bethe \cite{be} improved upon the
Bragg-williams approximation by treating somewhat more
accurately the interaction between the nearest neighbours. It
turns out that this approximation is infact exact for
1-dimension and hence does not predict phase transition at any
finite nonzero T while for 2 and higher dimensions, it
does predict spontaneous magnetization. Soon
afterwords, Peierls \cite{pe} wrote a paper entitled ``On Ising's
Model of Ferromagnetism" in which he gave a simple argument
showing that at sufficiently low temperature, Ising model in 2
or 3 dimensions must exhibit spontaneous magnetization. It
turned out much later that his argument involved an incorrect
step \cite{fs} but nevertheless, the conclusion and the general
procedure are correct.  This was an important step, and the
method is used still now in various situations.

The first exact quantitative result for the 2-dimensional Ising
model was obtained by Kramers and Wannier \cite{kw} when they
located the transition temperature by using the symmetry of the
two-dimensional lattice to relate the high- and low-$T$
expansion of the partition function. On February 28, 1942 there
was a meeting of the New York academy of Sciences in which
Wannier gave a talk on this work. At the end of his talk Onsager
made a remark announcing that he has been able to obtain the
exact solution of the 2-dimensional Ising model for square
lattice and without external magnetic field. It is very
significant though that he published his results in a journal
only two years later. Infact this was always his style.  He
could see through the math and get the final result but was
careful to publish his results only after he has satisfactorily
settled all the mathematical questions.

\section{ONSAGER$-$HIS LIFE AND WORKS}

Lars Onsager was born at Oslo in 1903. Even though he got his
degree in chemical engineering in 1925 he devoted most of his
time to studying math and physical sciences. In 1931 he wrote
two monumental papers in Physical Review about reciprocal
relations in irreversible processes. This essentially started
the new branch of thermodynamics $-$ that of irreversible
processes. Eventually, for this work, he received Nobel prize in
Chemistry in the year 1968. The Nobel citation noted that the
publication (two papers in Physical Review of length 22 and 15
pages) was one of the smallests ever to be awarded a Nobel prize!
In 1933 he was offered the Sterling Post-Doctoral fellowship at
the Chemistry Department of Yale University. But there was one
small problem. Onsager had never written his Ph.D. thesis! He
was persuaded to write one which he did on the properties of
Mathieu functions. This work is recognised as one of the
significant contributions to the subject. In 1934 he became the
faculty member of the department. From 1944 till his retirement
in 1972 he occupied the J. Willard Gibbs chair of Theoretical
Chemistry at Yale University. In 1972, he moved to Miami
University where he worked in the Center for Theoretical Studies
till his death in 1976.

During his entire career, Onsager only published about 60 papers
but each was an important one published only after he was
fully satisfied with its significance and its mathematical
rigour. Many a times he announced his results in conferences as
a remark after some talk but published the results much later or
never did! We will see one such example below. He would call
himself a chemist but truely speaking he was one of the last
truely universalist of this century.
For more information see Ref. \cite{fishons}.

\section{ONSAGER'S EXACT SOLUTION OF 2-D ISING MODEL}

To appreciate Onsager's exact solution, it is worth recalling
the work of Kramers and Wannier \cite{kw} before Onsager where
they showed that the partition function for this problem can be
written as the largest eigenvalue of a certain matrix. Onsager's
method was similar to these authors except that he emphasized
the abstract properties of rather simple operators rather than
their explicit representation by unwieldy matrices. It must be
said here that his method is highly nontrivial and complicated.
Actually his derivation can be easily followed step by step but
the over-all plan is quite obscure. Since a simplified
version as given by Bruria Kaufman \cite{ka} has
been discussed at a number of places including textbooks we
shall merely quote the result. Onsager showed that the canonical
partition function $Q (B=0,T)$ in the limit $N \rightarrow
\infty$ is given by
\be \label{eq4}
\lim_{N\rightarrow\infty} \ln Q  ( B = O,T) = \ln (2 \cosh
(2\beta J)) + {1\over 2\pi}\int_0^{\pi} \ d\phi\ \ln \ {1\over
2}(1+\sqrt{1-\kappa^2 \sin^2\phi}),
\ee
where $\kappa \equiv {2\sinh{(2\beta J)}/\cosh^2{(2\beta J)}}$. From here
it followed that the specific heat $C (B=0,T)$ defined by
\be \label{eq5}
C (B,T) =  k\beta^2 {\partial^2\over\partial\beta^2} [\ln Q
(B,T)],
\ee
diverges logarithmically as $T \rightarrow T_c$ where $T_c$ is
given by
\be \label{eq6}
\tanh {2J\over k T_c} = {1\over\sqrt 2} \Longrightarrow {k
T_c\over J} = 2.269185.
\ee
For comparison, $kT_c/J$ is predicted to be 4 and 2.88 in
the mean field and Bethe approximations respectively.

Over the years, several alternative simplified methods have been
given but all of them are quite involved and lengthy.
Bruria Kaufman \cite{ka} gave a simplified proof which
is based on spinorial representation of the rotation group and
it is this proof which is essentially given in K. Huang's book
on statistical mechanics. Further, using Onsager's proof, the
exact partition function and hence other properties of several
other 2-dimensional lattices have been deduced. Finally, Onsager
and Bruria Kaufman \cite{ko} calculated spin
correlation functions.

{\bf Further Developments:} To justify calling the phenomemon at
$T = T_c$ a phase transition, one has to examine the long range
order i.e. spontaneous magnetization and show that it is
nonzero. This is a very difficult calculation since it has to be
done with $B \neq 0$ (but could be small) at the beginning of
the calculation and finally putting it equal to 0 at the end of
the calculation. On August 23, 1948 there was a talk by Tisza at
Cornell University. At the end of the talk, Onsager walked up to
the blackboard and coolly announced that he and
Bruria Kaufman had solved this problem and they indeed found
that the long range order is nonzero.  Infact he even wrote down
the formula on the blackboard. He repeated his comments at the
first postwar IUPAP conference on Statistical Mechanics at
Florence in 1949 after a talk by Rushbrook. However, Kaufman and
Onsager never published their calculatuions ; it only appeared
as a discussion remark \cite{on1}. In print the first full
calculation was infact published by Yang \cite{ya}. Why did
Onsager not publish his results? Years later, only in 1969 he
gave the reason.  In computing the long range order, Onsager was
led to a general consideration of Toeplitz matrices but he did
not know how to fill out holes in the maths$-$by the time he
did, mathematicians were already there! What Onsager and Yang
had obtained was that as $T$ goes to $T_c$ from below then the
long range order $\overline L(B=0,T)$ is given by
\be \label{eq7}
\overline L(B=0, T) = (1 - {T/T_c})^{1/8}
\ee
while it is zero if $T$ approaches $T_c$ from above.  It may be
noted here that both the mean field and the Bethe approximation
predict completely wrong behaviour i.e. they predict that
$\overline L$ will go like $(1-{T/ T_{c}})^{1/2}$.  These
approximations do not also reproduce the logarithmic divergence
of the specific heat.  This difference with the approximate
theory is rather important. One realizes that the nature of singularity
is rather subtle.  No matter what approximation one does, short
of an exact solution, one never gets the log.  This drives home
a point that the singularity is due to
certain special circumstances that get killed by the approximations.
It was several years later, when it could be identified as the effect
of fluctuation, as reflected through fluctuation - response
theorem. The universality of various phase transitons comes from
the nature of fluctuations. It is the renormalization group
approach developed by Kadanoff, Wilson and others, that gave the
proper framework to handle fluctuations \cite{domb}.

{\bf Unsolved Problems:} To appreciate how nontrivial Onsager's
exact solution of the 2-dimensional Ising model was, let us
remember that till today the exact solution of the 2-dimensional
Ising model in nonzero external magnetic field has not been
obtained. Further, the 3-dimensional Ising model with or without
external magnetic field is also an unsolved problem. Similarly
the 2-dimensional Ising model with any additional interaction (as,
e.g., nearest and next-to nearest
neighbour interactions) is also an unsolved problem.  However,
special cases like the three spin interaction on a traingular
latice can be solved \cite{bax}.

\section{INFLUENCE}

Let us go back to the paper and discuss some of the developments
where the 1944 paper played a direct role.

(1) {\bf Duality }: Duality between high temperature and low temperature
for the Ising partition function was known before the Onsager
Solution. However, Onsager used it to the full extent and introduced
the idea of star-triangle (ST) transformation that could be used to solve
the triangular lattice problem. The existence of duality is important
because ``it always converts order into disorder and vice versa''\cite{on}.
Since the low temperature ordered phase is characterized by an order
parameter, it seems possible to define an analogous quantity for the
high temperature phase also. It is only in the eighties that such
``disorder'' parameters played an important role in conformal
invariance. The ST transformation turned out to be important in a
different development when attempts were made in the 70's to solve
more complex models, like vertex models. The existence of duality and
ST guarantees the existence of commuting transfer matrices as
expressed through the Yang-Baxter equation. The solvability of these
models relies  heavily  on such commuting matrices \cite{bax}.

(2) {\bf Transfer matrix }: The major problem was to diagonalize the
transfer matrix. The largest eigenvalue determines the thermodynamic
behaviour. Onsager could find all the eigenvalues for a finite
lattice. This had implications far beyond the original idea of
getting only the free energy.

Onsager's result showed that for a square (or any two dimensional)
lattice, four different terms are needed if periodic boundary
conditions are used. Why four was answered in a clear way by Kasteleyn
in 1963 when he gave a general procedure for solving the Ising model
using combinatorics (``dimers''). Kasteleyn showed that for a surface
of genus g, $4^g$ ``terms'' are needed. For open boundary conditions
i.e. for strictly planar lattices , g = 0 while, for periodic
boundary conditions, g = 1 as it has the geometry of a torus \cite{kast}.
This also indicates
that pushing the same approach to solve a 3-d lattice would require
an infinite number of terms.

(3) {\bf Correlation length and correlation}:

Since Onsager could find all the
eigenvalues, he could also identify the correlation length. If
$\lambda_1 >\lambda_2 > \lambda_3$... are the eigenvalues, then the
partition function can be written as
\be \label{eq:part}
Q = \lambda_1^N \left [ 1 + \left
({\lambda_2\over\lambda_1}\right )^{^N}+....\right ],
\ee
where N is the number of rows in the direction of transfer. The phase
transition occurs when the largest eigenvalue is degenerate. Close to
the critical point, one can write $(\lambda_2/\lambda_1)^N \sim
e^{-N/\xi}$ where $\xi = - (\mid \ln \lambda_2/\lambda_1\mid )^{-1}$.
This identifies $\xi$ as a special length scale to characterize the
critical behaviour.  The degeneracy of the eigenvalues as $T
\rightarrow T_c$, in turn, implies a diverging length scale at
the transition point. In this way, Onsager showed that $\xi \sim \mid
T-T_c\mid ^{-\nu}$ with $\nu=1$. The renormalization group approach mentioned
earlier is based on the idea of a diverging length scale so
that the system is scale invariant at the
critical point.

Later on, in 1949 Kaufman and Onsager showed
that at $ T= T_c$, spin correlation $\Gamma (r) \equiv < S(0) S(\vec
r) >$ deccays as $r^{-1/2}$ for
large distance\cite{ko}. In 1965, Patashinskii-Pokrovsky suggested the now
famous scaling form $\Gamma (r,T)\approx r^{-1/4} D(r/\xi)$ where,
apart from the power law prefactor, all distances are scaled by the
correlation length at that temperature \cite{pata}. Explicit calculations could
put the correlation function, in the proper asymptotic limit, in the
scaling form but it was also noted that for $ T < T_c, \Gamma \sim
e^{-r/\xi}/r^2$. This  anomalous behaviour in the low temperature
phase is a very peculiar property of the Ising model in zero magnetic
field. It is in the eighties that this point got clarified through the use
of directed polymers \cite{fisher}.

(4) {\bf Boundary effects} : Onsager recognized the importance of boundary
conditions. He realized that by changing the couplings in one row to
antiferromagnetic coupling $(J\rightarrow -J)$ would increase the
free energy in the ordered state because across this line spins
cannot be aligned as in the bulk. This change, in other words, will
force an interface in the system. The increase in free energy is then
the energy of the interface or the surface tension. Since the
interfaces vanishes at the critical point, the surface tension also
vanishes at $T = T_c$. The exact solution showed that the
surface tension vanishes as $(T_c-T)^{\mu}$ with $\mu=1$. More
general arguments later on
proved that $\mu +\nu = 2$ for the two dimensional Ising model.
For any other model the right hand side would involve the
specific heat exponent also.

(5) {\bf Finite size effects}: Since the partition function was known for
finite lattices, the specific heat could also be calculated. It was shown
that, at $T = T_c$, the specific heat per spin, $C$, is finite. No
phase transition
can take place in a finite system. Furthermore, $C \sim \ln N$ as
$N\rightarrow \infty$ (for a $N\times\infty$ lattice). It took many
years to grasp the significance of this result until the idea of
finite size scaling was introduced in 1970. Finite size scaling now
plays an important role in analysis of numerical data in wide
varities of problems.

A more dramatic result, though cannot be found explicitly, in the
Onsager paper, is the finite size behaviour of the free energy.
Collecting various terms, from the calculation of the speciofic heat,
it was found that for an $m\times n$ lattice
\be
\lim_{m\rightarrow\infty} {ln Z_{mn}(T_c)\over mn} =
{2G\over\pi} +{1\over 2} \ln 2 + {\pi\over 12} {1\over n}
+..  ,
\ee
where $G$ is the Catalan constant.  This is to be contrasted
with Eq. \ref{eq:part} for $T\neq T_c$ where there is an
exponential approach to the large lattice (thermodynamic) limit .
In 1986, Affleck argued using conformal invariance that in two
dimensions, at criticality,
$$\lim_{m\rightarrow\infty} {ln Z\over mn} = const +{\Lambda c\over 6
r}$$
with $c$ the central charge as the only quantity required to identify the
critical behaviour (`universality class') in 2 dimensions \cite{aff}. A
comparision of the two gives c = 1/2 for the Ising model.

(6) {\bf Experiment}: The
simplicity and wide applicability of the two dimensional Ising model
and its solution sill leave behind a queer sensation that, after
all, the model is a bit artificial.  So far as the exponents are
concerned, they are universal and a wide variety of systems do
have Ising exponents.  It however remained a challenge to
find an experimental system that could exhibit the behaviour one
finds from a landmark of theoretical physics.
The challenge was met when an alloy was found in 1974 which
showed the behaviour of
magnetization exactly like the Onsager result with only $T_c$ as the
adjustable parameter \cite{exper}.

\end{document}